\pdfoutput=1
\RequirePackage[T1]{fontenc}
\RequirePackage{fix-cm}
\RequirePackage{lmodern}
\documentclass[a4paper,12pt,pdftex]{article}
\usepackage[pass]{geometry}

\usepackage{graphicx}
\newcommand{\ket}[1]{|#1 \rangle}

\newtheorem{theorem}{Theorem}
\newtheorem{proposition}[theorem]{Proposition}

\begin{document}
\title{Lee Weight for Nonbinary Quantum Error Correction}
\author{Ryutaroh Matsumoto,\\
Tokyo Institute of Technology, Japan.}
\date{26 May 2021}
\maketitle
\begin{abstract}
  We propose the quantum Lee weight for quantum errors,
  provide a Gilbert-Varshamov type bound,
  and a code construction for the proposed weight.
\end{abstract}

\section{Introduction}
Let $\ell \geq 2$ be an integer,
and $\mathbf{Z}_\ell = \{ 0$, $1$,
\ldots, $\ell-1\}$ be the ring of integers modulo $\ell$.
The Lee weight $w_L(x)$ for $x \in \mathbf{Z}_\ell$
is $x$ if $0\leq x \leq \ell/2$ and $\ell - x$ if $\ell/2 \leq x \leq \ell - 1$
\cite{berlekampbook}.
The Lee weight of a vector $\vec{x} = (x_1$,
\ldots, $x_n) \in \mathbf{Z}_\ell^n$
is defined by $w_L(\vec{x}) = w_L(x_1) + \cdots
+ w_L(x_n)$.
In the conventional digital communications,
when symbols in $\mathbf{Z}_\ell$ are modulated by
$\ell$-PSK, lower Lee weight of an error vector $\vec{x}$ roughly corresponds to
the higher probability of the additive noise corresponding to $\vec{x}$
\cite{berlekampbook}.
Therefore, for the same information rate, Lee weight can provide
better error probabilities for the conventional digital communications.

We turn our attention to the quantum errors.
Suppose that we have an $\ell$-dimensional quantum system
with an orthonormal basis $\{ \ket{0}$, \ldots, $\ket{\ell-1}\}$
\cite{chuangnielsen}.
Consider quantum errors $X \ket{i} = \ket{i \bmod \ell}$
and $Z \ket{i} = \lambda^i \ket{i}$, where $\lambda$ is a primitive $\ell$-th
root of $1$.
Consider a quantum error $M(a,b) = \lambda^i X^a Z^b$ \cite{ashikhmin00},
where $a,b \in \mathbf{Z}_\ell$.
Usually it has been assumed that the probabilities
of $M(a,b)$ and $M(a',b')$ are the same regardless of
$(a,b)$ and $(a',b')$ provided that both $(a,b)$ and $(a',b')$ are nonzero vectors.
On the other hand, like $\ell$-ASK in the conventional digital modulation,
under some situation the probability of $M(a,b)$
is more closely approximated as \cite{otten20}
$\gamma \times \alpha^{w_L(a)} \beta^{w_L(b)}$, where $0^0$ is $1$ and
$0 \leq \alpha, \beta < 1$ and $\gamma$ the normalizing constant
so that the sum of all probabilities is $1$.
Under such a situation,
the probability of a quantum error $\lambda^i X(\vec{a})Z(\vec{b})$
is \emph{proportional} to $\alpha^{w_L(\vec{a})} \beta^{w_L(\vec{b})}$.
Therefore, use of the Lee weight is more consistent with the probabilities of
corresponding errors.
Another reason for using the Lee weight is higher information rate can
be realized by codes for the Lee weight than codes with the conventional quantum
distance of the same weight. This fact will be demonstrated in Section \ref{sec:comparison}.

This paper is organized as follows:
Section \ref{sec2} introduces notations.
Section \ref{sec3} proposes a Gilbert-Varshamov type bound.
Section \ref{sec4} proposes a code construction based on \cite{berlekampbook}.

\section{Preliminaries}\label{sec2}

In the following, we restrict our attention to the CSS-type QECC
where the quantum bit errors and the quantum phase errors are
independently identified and corrected.
In such a case, larger fidelity of error correction
is provided by regarding the bit and the phase errors with lower Lee weights as
more likely, and to choose the errors of the lowest Lee weight
as the error to be corrected.
Therefore, we define the quantum Lee weights of $\lambda^i X(\vec{a})Z(\vec{b})$
as the pair $(w_L(\vec{a}), w_L(\vec{b}))$.
A QECC is defined to have the minimum quantum Lee weights
$(d_X, d_Z)$ if it can detect every error with Lee weights $(\delta_X$,
$\delta_Z)$ whenever $\delta_X < d_x$ and $\delta_Z < d_Z$.

Hereafter we restrict $\ell$ to be some prime integer $p$.
We consider the quantum stabilizer code defined by the
stabilizer $C_1 \times C_2 \subset \mathbf{F}_p^{2n}$,
with $C_1 \subset C_2^\perp$ and $C_2 \subset C_1^\perp$,
where $C_i^\perp$ denotes the usual dual code
of $C_i$ with respect to the standard Euclidean inner product.

\section{Gilbert-Varshamov bound}
\label{sec3}
We devote this section to provide a finite and an asymptotic Gilbert-Varshamov-type (GV) bound for quantum codes for the quantum Lee weights. We start with the finite case.
\subsection{Finite GV bound}
Let $N(p, n,d)$ be the number of vectors $\vec{x} \in \mathbf{F}_p^n$
with $w_L(\vec{x}) \leq d$.

\begin{proposition}\label{prop1} \cite[Proposition 10.10]{rothbook}
If $d < p/2$, then
\[
N(p,n,d) = \sum_{i=0}^n 2^i {n \choose i} {d \choose i}
\]
\end{proposition}

\begin{proposition}\label{prop2}
\[
N(p,n,d) \leq
{2 n +d - 1 \choose d} + {2 n + d - 2\choose d-1}
\]
\end{proposition}
{\textbf{Proof.}}
There are $n$ positions in $\vec{0} = (0$, \ldots, $0) \in \mathbf{F}_p^n$.
One can add $+1$ or $-1$ to the $i$-th zero in $\vec{0}$.
If both $+1$ and $-1$ are added to the same coordinate, they cancel each other.
If one adds $+1$ or $-1$ $d$ times, then the resulting vector has
Lee weight $\leq d$. When no $+1$ or $-1$ are canceled,
the resulting Lee weight is $d$. If one $+1$ is canceled,
there must exist another $-1$ that cancels $+1$.
Therefore the number of cancellations is an even integer.
If one adds $+1$ or $-1$ $d$ times, then the resulting Lee weight is
$d$, $d-2$, $d-4$, \ldots.
All the vector with Lee weights $d$, $d-2$, \ldots, can be constructed
in the above way.
The number of ways of adding $+1$ or $-1$ $d$ times is
at most
the number of multisets of cardinality $d$, with elements taken from a finite set
of cardinality $2n$, which is
\[
{2n + d -1 \choose d}.
\]
By a similar argument, we can also see that
the number of vectors whose Lee weights are $d-1$, $d-3$, \ldots is
upper bounded by
\[
{2 n + d - 2 \choose d-1}.
\]
\hfill\rule{1ex}{1ex}


Let us start with our result.
\begin{theorem}
\label{FGV}
Consider positive integer numbers $n, k_1, k_2$, $d_x$ and $d_z$ such that $k_1 \leq n$, $k_2 \leq n$
which satisfy the following inequality

$$
\frac{p^{n-k_1} - q^{k_2}}{p^n-1} N(p,n,d_z-1)   +
\frac{p^{n-k_2} - q^{k_1}}{p^n-1} N(p,n,d_x-1) < 1,
$$
then there exists an $[[n,n-k_1-k_2]]_q$ QECC whose
minimum quantum Lee weights are at least  $(d_x, d_z)$.
\end{theorem}
\noindent\textbf{Proof:}
For simplicity sake, in this proof $C'_2$ will be used instead of $C_2^\perp$.
Consider integer numbers $n,k_1,k_2$  as in the statement. Define
\begin{eqnarray*}
&&A(n,k_1,k_2) =\big\{ (C_1, C'_2) \; | \;
C_1, C'_2 \subset \mathbf{F}_q^n, \\
&&\dim C_1 = k_1, \dim C'_2 = n-k_2, \mbox{and } 
C_1 \subset C'_2 \big\}.
\end{eqnarray*}
For $\mathbf{v} \in \mathbf{F}_p^n$, define also
$$
B_{z}(\mathbf{v}) = \big\{ (C_1, C'_2) \in A(n,k_1,k_2) \; | \; 
\mathbf{v} \in C_1^\perp \setminus (C^{\prime\perp}_2 \cap C_1^\perp) \big\}
$$
and
$$
B_{x}(\mathbf{v}) = \big\{ (C_1, C'_2) \in A(n,k_1,k_2) \; | \; 
\mathbf{v} \in C'_2 \setminus (C_1 \cap C'_2) \big\}.
$$
For nonzero $\mathbf{v}_1$ and $\mathbf{v}_2 \in \mathbf{F}_q^n$,
we claim that
$$\# B_{z}(\mathbf{v}_1)=\# B_{z}(\mathbf{v}_2),$$ where we recall that $\#$ means cardinality.

Let us see a proof. Denote by $GL(n,q)$ the set of invertible matrices on $\mathbf{F}_p^n$ and for a fixed $(D_1, D'_2) \in A(n,k_1,k_2)$,
a fixed $M_1 \in GL(n,q)$ with $M_1 \mathbf{v}_1 = \mathbf{v}_2$
and $M'_1  \in GL(n,q)$ with $M'_1 \mathrm{span}(\mathbf{v}_1)^\perp
= \mathrm{span}(\mathbf{v}_2)^\perp$, where $M'_1 \mathrm{span}(\mathbf{v}_1)^\perp$ stands for the linear space given by the products $M'_1 \mathbf{w}$ such that $\mathbf{w} \in \mathrm{span}(\mathbf{v}_1)^\perp$. Then we have
\begin{eqnarray*}
  && \# B_{z}(\mathbf{v}_1)\\
  &=& \# \big\{ (C_1, C'_2) \in A(n,k_1,k_2) |  \mathbf{v}_1 \in C_1^\perp \setminus (C^{\prime\perp}_2 \cap C_1^\perp)\big\}\\
  &=& \# \big\{ (C_1, C'_2) \in A(n,k_1,k_2) |
\mathrm{span}(\mathbf{v}_1)^\perp \supseteq C_1   \mbox{ and } \mathrm{span}(\mathbf{v}_1)^\perp
\not\supseteq C'_2 \big\}\\
  &=& \# \big\{  (M D_1, MD'_2) \;| \;
\mathrm{span}(\mathbf{v}_1)^\perp \supseteq MD_1   \mbox{ and } \mathrm{span}(\mathbf{v}_1)^\perp
\not\supseteq MD'_2, M \in GL(n,q) \big\}\\
  &=& \# \big\{  (M'_1M D_1, M'_1MD'_2)|
M'_1 \mathrm{span}(\mathbf{v}_1)^\perp  \supseteq \! M'_1MD_1 \\
&& \mbox{ and } M'_1 \mathrm{span}(\mathbf{v}_1)^\perp
\not\supseteq M'_1MD'_2, M \in GL(n,q) \big\}\\
  &=& \# \big\{  (M'_1M D_1, M'_1MD'_2) |
\mathrm{span}(\mathbf{v}_2)^\perp \supseteq M'_1M D_1 \\
&& \mbox{ and } \mathrm{span}(\mathbf{v}_2)^\perp \not\supseteq M'_1MD'_2, M'_1M \in GL(n,q) \big\}\\
&=& \# B_{z}(\mathbf{v}_2).
\end{eqnarray*}
We also claim that
$\# B_{x}(\mathbf{v}_1)=\# B_{x}(\mathbf{v}_2)$. Indeed,
\begin{eqnarray*}
  && \# B_{x}(\mathbf{v}_1)\\
  &=& \# \big\{ (C_1, C'_2) \in A(n,k_1,k_2) |
\mathbf{v}_1 \in C'_2 \setminus (C^{\prime}_2 \cap C_1) \big\}\\
  &=& \# \big\{ (MD_1, MD'_2)  \;| \;
\mathbf{v}_1 \in MD'_2 \setminus (MD^{\prime}_2 \cap MD_1),   M \in GL(n,q) \big\}\\
  &=& \# \big\{ (M_1MD_1, M_1MD'_2) |  
M_1\mathbf{v}_1 \in M_1MD'_2 \setminus (M_1MD^{\prime}_2 \cap M_1MD_1),\\&& M_1M \in GL(n,q) \big\}\\
  &=& \# \big\{ (MD_1, MD'_2)  \;| \;
M_1\mathbf{v}_1 \in MD'_2 \setminus (MD^{\prime}_2 \cap MD_1),   M \in GL(n,q) \big\}\\
&=& \# B_{x}(\mathbf{v}_2).
\end{eqnarray*}

Next we will count the quantity of triples $(\mathbf{v}$, $C_1$, $C'_2)$
such that $\mathbf{v} \in C_1^\perp \setminus (C^{\prime\perp}_2 \cap C_1^\perp)$
in two different ways.
We observe that
$$
\dim C^{\prime\perp}_2 \cap C_1^\perp
= \dim C_2 \cap C_1^\perp \\=
\dim C_2 - (\dim C_2 - \dim C_2 \cap C_1^\perp) = k_2 
$$
For each pair $(C_1, C'_2) \in A(n,k_1,k_2,0)$ there are
\[
q^{n-k_1} - q^{k_2 - 0}
\]
vectors $\mathbf{v}$ such that $\mathbf{v} \in C_1^\perp \setminus (C^{\prime\perp}_2 \cap C_1^\perp)$.
Thus the total number of such triples is
\[
(q^{n-k_1} - q^{k_2 }) \# A(n,k_1,k_2).
\]

On the other hand, we can count the total number of triples as
\[
\sum_{\mathbf{0}\neq \mathbf{w} \in \mathbf{F}_q^n}
\# B_{z}(\mathbf{w}) = (q^n-1) \# B_{z}(\mathbf{v})
\]
for any fixed nonzero $\mathbf{v}$.
This implies
\[
\frac{\# B_{z}(\mathbf{v})}{\# A(n,k_1,k_2)}
= \frac{q^{n-k_1} - q^{k_2 }}{q^n-1}.
\]

A similar argument shows
\[
\frac{\# B_{x}(\mathbf{v})}{\# A(n,k_1,k_2)}
= \frac{q^{n-k_2} - q^{k_1 }}{q^n-1}.
\]

If we remove a pair $(C_1, C'_2)$ from
$A(n,k_1,k_2)$ either when
$\mathbf{v}_z \in C_1^\perp \setminus (C^{\prime\perp}_2 \cap C_1^\perp)$ or
when $\mathbf{v}_x \in C^{\prime}_2 \setminus (C^{\prime}_2 \cap C_1)$
for $1 \leq w_L(\mathbf{v}_z) \leq d_z-1$ and
for $1 \leq w_L(\mathbf{v}_x) \leq d_x-1$, then  we remove in total
\begin{equation}
\sum_{1 \leq \mathrm{wt}(\mathbf{v}_z) \leq d_z-1} \# B_z(\mathbf{v}_z) +
\sum_{1 \leq \mathrm{wt}(\mathbf{v}_x) \leq d_x-1} \# B_x(\mathbf{v}_x) \label{eq1}
\end{equation}
pairs from $A(n,k_1,k_2)$.

As a consequence, there exists at least one
desired code whenever the number (\ref{eq1}) is less than $\# A(n,k_1,k_2)$ which proves the statement.
\hfill\rule{1ex}{1ex}

\subsection{The asymptotic GV bound}
In order to derive an asymptotic version,
we will derive an asymptotic form of Proposition \ref{prop2}.
Set
$h_p(y) :=  -y \log_p y -(1- y) \log_p (1-y)$ the $p$-ary entropy function.
We have
\begin{eqnarray}
&&{2 n+d-1 \choose d} +{2 n+d-2 \choose d-1}\nonumber\\
&\leq&
{2 n+d-1 \choose d} +{2 n+d-1 \choose d-1}\nonumber\\
&\leq& \exp_2 ( n(2+d/n) h_2(d / n(2+d/n+1/n)))\\
&=& \exp_p (n(2+d/n) h_p(d / n(2+d/n+1/n))). \nonumber
\end{eqnarray}

By using the above bound, we can derive the following
asymptotic sufficient condition of existence of quantum codes.

\begin{theorem}\label{thm3}
Consider positive real numbers $K_1, K_2, \delta$ such that
\begin{eqnarray*}
n(2+\delta_z) h_p(\delta_z / n(2+\delta_z))  &<& K_1,\\
n(2+\delta_x) h_p(\delta_x / n(2+\delta_x))  &<& K_2.
\end{eqnarray*}
Then, for sufficiently large $n$, there exists an asymmetric EAQECC with parameters
$\big[\big[n, \lfloor n-nK_1-nK_2+n\lambda\rfloor  \big]\big]_q$
with Lee weights at least $(\lfloor n\delta_z\rfloor, \lfloor n\delta_x\rfloor)$. \rule{1ex}{1ex}
\end{theorem}

\subsection{Comparison with the Hamming weight}\label{sec:comparison}
\begin{figure}
\includegraphics{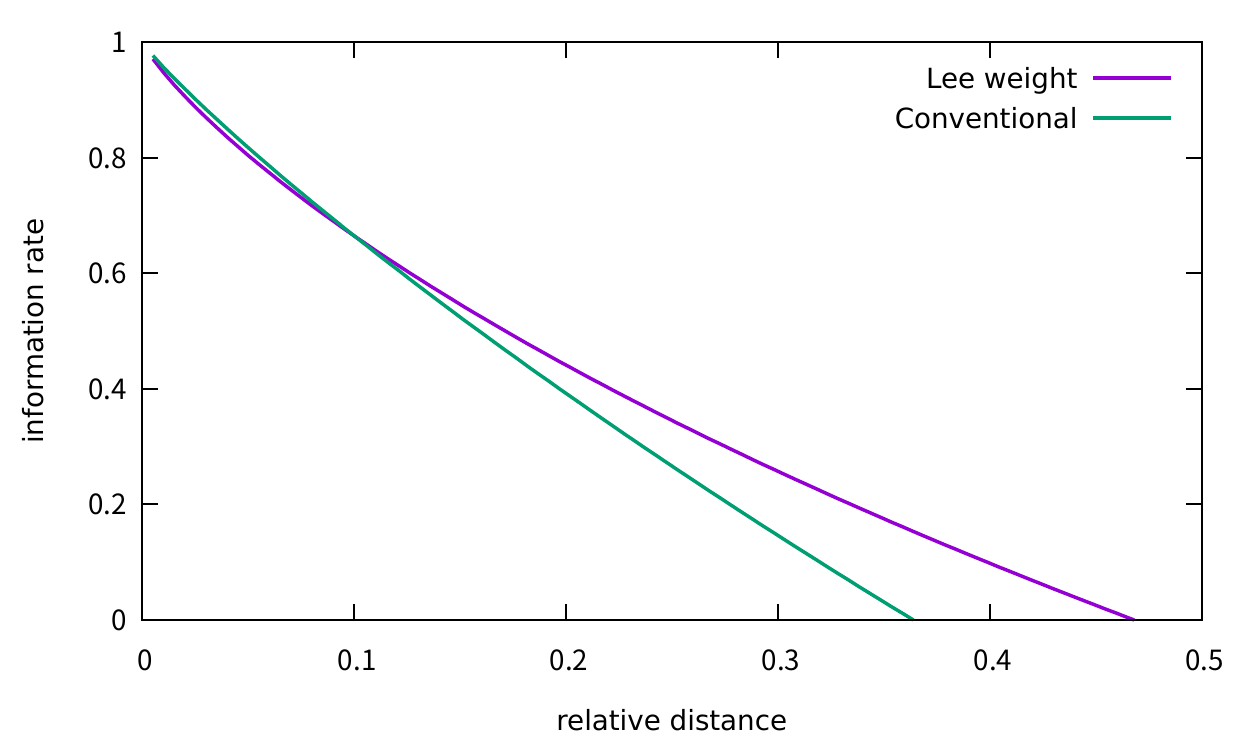}
\caption{Information rates for correcting errors less than relative weight $\delta$}\label{fig1}
\end{figure}

We will compare the information rates for for detecting
errors less than $n\delta$ weight and $n\delta$
Lee weights for asymptotically large $n$,
which corresponds to $\delta = \delta_x = \delta_z$
in Theorem \ref{thm3}.
By \cite{feng04,galindo19,jin11}, the information rate of the conventional
quantum weight is $R = 1 - h_p(\delta) - \delta \log_p (p^2-1)$,
while by Theorem \ref{thm3}
the information rate $R_L$ for Lee weight is 
$1-2(2+\delta) h_p(\delta / n(2+\delta))$.
Those information rates are compared in Figure \ref{fig1}
for $p=11$. We can see that Lee weight can provide higher information rates.

\section{Code Construction}\label{sec4}
We propose a code construction for a special case $C = C_1 = C_2$.
Suppose that we have a quantum error $(\vec{e}_x|\vec{e}_z) \in \mathbf{F}_p^{2n}$,
and that $\vec{h}_1$, \ldots, $\vec{h}_r$ span $C$ as a linear space.
The standard measurement procedure \cite{chuangnielsen}
provides
$\langle \vec{e}_x$, $\vec{h}_i\rangle$ and
$\langle \vec{e}_z$, $\vec{h}_i\rangle$ for $i=1$, \ldots, $r$.

A quantum decoder has to choose
argmin $\{ w_L(\vec{e'}_x) \mid
\langle \vec{e'}_x$, $\vec{h}_i\rangle = \langle \vec{e}_x$, $\vec{h}_i\rangle$
for $i=1$, \ldots, $r\}$, and also
argmin $\{ w_L(\vec{e'}_z) \mid
\langle \vec{e'}_z$, $\vec{h}_i\rangle = \langle \vec{e}_z$, $\vec{h}_i\rangle$
for $i=1$, \ldots, $r\}$.
The above steps are exactly the same as the standard decoding procedure
of classical linear codes with respect to the Lee weight.
Thus, we will show a code construction $C \subseteq C^\perp$
such that $C^\perp$ has a decoding procedure with respect to the
Lee weight.

We will use the construction of negacyclic codes in \cite{berlekampbook}.
In the following,
we assume the code length $n$ satisfies $p \not | 2n$ so that
negacyclic codes of length $n$ can exist.
Let $\alpha$ be a primitive $2n$-th root of unity over $\mathbf{F}_p$.
We choose $C^\perp \supset C$ as negacyclic codes.
Let $g(x)$ be the generator polynomial of $C^\perp$,
and $h(x)$ be that of $C$.
We have $h(x) = \hat{g}(0)^{-1} x^{\deg \hat{g}(x)} \hat{g}(1/x)$,
where $\hat{g} = (x^n+1)/g(x)$  \cite{rothbook}.

We choose $g(x)$ such that
$g(\alpha) = g(\alpha^3) = \cdots =
g(\alpha^{2t-1})=0$ so that the minimum Lee weight of
$C^\perp \geq \min\{p, 2t+1\}$ \cite{berlekampbook}.
We need $C^\perp \supseteq C$, which is equivalent to
$g(x)|h(x)$,
so that we can apply the quantum CSS construction on $C$.

By summarizing the above observation,
we have the following theorem:
\begin{theorem}\label{thm1}
Let $\alpha$ be a primitive $2n$-th root of unity over $\mathbf{F}_p$.
Let $g(x)$ is a factor of $x^n+1$,
satisfying $g(\alpha) = g(\alpha^3) = \cdots =
g(\alpha^{2t-1})=0$ for $ \geq 1$.
The negacyclic code $C^\perp$ defined by $g(x)$
has length $n$ and the minimum Lee weight at least $\min\{p, 2t+1\}$.
We further assume $g(x)$ divides $h(x)$, where $h(x)$ is as defined above,
which ensures $C^\perp \supseteq C$.
By applying the quantum CSS construction on $C^\perp \supseteq C$,
we obtain an $[[n,n-2 \dim C]]_p$
quantum code. By using the standard decoding procedure for negacyclic
codes in \cite{berlekampbook}, the quantum bit and phase errors can
be corrected up to $t$ Lee weight. \rule{1ex}{1ex}
\end{theorem}


\end{document}